\documentclass[usenatbib,usegraphicx]{mn2e}
\usepackage{times,amsmath,amssymb,afterpage}
\begin{document}

\title{Short-period variability in the Class II methanol maser source G12.89+0.49 (IRAS 18089-1732)}

\author[Goedhart et al.]
{S. Goedhart$^{1}$\thanks{E-mail: sharmila@hartrao.ac.za}, M. C. Langa$^{1,2}$, M. J. Gaylard$^{1}$ and
D. J. van der Walt$^{2}$ \\ 
$^1$ Hartebeesthoek Radio Astronomy Observatory, PO Box 443,
Krugersdorp, 1740, South Africa\\ 
$^2$     School of Physics,
North-West University, Potchefstroom campus, Private Bag X6001, Potchefstroom, 2520, South Africa \\
}

\date{Accepted ;      Received ;      in original form }

\pagerange{\pageref{firstpage}--\pageref{fig:walsh}}
\pubyear{2008}

\maketitle

\begin{abstract}
Time series are presented for the class II methanol maser source G12.89+0.49, which has been monitored for nine years at the Hartebeesthoek Radio Astronomy Observatory. The 12.2 and 6.7 GHz methanol masers were seen to exhibit rapid, correlated variations on timescales of less than a month. Daily monitoring has revealed that the variations have a periodic component with a period of 29.5 days. The period seems to be stable over the 110 cycles spanned by the time series.  There are variations from cycle to cycle, with the peak of the flare occurring anywhere within an eleven day window but the minima occur at the same phase of the cycle. Time delays of up to 5.7 days are seen between spectral features at 6.7 GHz and a delay of 1.1 day is seen between the dominant 12.2 GHz spectral feature and its 6.7 GHz counterpart.  
\end{abstract}

\begin{keywords} 
masers -- HII regions -- ISM: clouds -- Radio lines: ISM -- stars: formation
\end{keywords}

\label{firstpage}

\section{Introduction}

Currently, understanding of the process of high mass star formation is limited by observational challenges.  Most regions of active high mass star formation are distant so observations need to be more sensitive and at higher resolution than for nearby low mass regions.  In addition the young objects are deeply embedded with thermal emission only readily detectable in the mid-infrared to submm range.  Class II methanol masers have proven to be a reliable tracer of early stages of massive star formation \citep{2006ApJ...638..241E}.  In many cases, the methanol masers are the only cm-wave emission associated with a young stellar object and are thus the only probe of conditions in the region. Since the 6.7 and 12.2 GHz transitions are strong, they can be readily observed with smaller telescopes. Monitoring of flux density variations in the masers can yield valuable information on changing conditions in the region and possibly the kinematics of the masers. However, since there is still uncertainty about the location of the masers in the environment of the massive young star, interpreting the variability can be challenging. High-resolution imaging shows that in some sources the masers have a linear distribution with velocity gradients, consistent with circumstellar discs \citep[e.g.][]{2005Ap&SS.295..231P} and \citet{2007A&A...464.1015V} find that, statistically, the maser velocity dispersions  are most consistent with a Keplerian disk. However, \citet{2009A&A...493..127D} argue against the disk hypothesis, suggesting that there may be a stronger link to outflows, and  \citet{2007A&A...461.1027G} infer that in AFGL 5142 the methanol masers may be associated with infalling gas.

The high incidence of variability in methanol masers is well established and was systematically studied at 6.7 GHz by \citet{2004MNRAS.355..553G}.  One of the sources in their sample, G12.89+0.49, showed rapid but undersampled variability. This methanol maser is associated with a high-mass protostellar object in IRAS 18089-1732, which has been observed in a number of different tracers. It is at a distance of 3.6 kpc with a bolometric luminosity of 10$^{4.5}$ L$_\odot$ \citep{2002ApJ...566..931S}. Hydroxyl masers have been detected at 1665 and 1667 MHz in left circular polarisation \citep{Cas98,2007A&A...465..865E} and 22 GHz water masers are also present \citep{Com90}. Methanol masers have been detected in the following transitions: 6.7 GHz \citep{Men91}, 12.2 GHz \citep{1988ApJ...331L..37K}, 44 GHz \citep{Sly94}, 95 GHz \citep{2000MNRAS.317..315V} and 107 GHz \citep{Cas00}. Unsuccessful searches were made at 9.978 GHz, 10.058 GHz \citep{1999A&A...343..939R} and 108 GHz \citep{Val99}.

An arcsecond-resolution map of the 6.7 GHz masers was produced by \citet{Wal98}. Submm observations with the SMA have shown an elongated structure with a velocity gradient perpendicular to a massive outflow \citep{2004ApJ...616L..23B,2005ApJ...628..800B,2008ApJ...673L..55B}. Only one lobe of the molecular outflow has been mapped, lying to the north of the core with a disc-like object perpendicular to it \citep{2004ApJ...616L..23B}. The outflow is believed to be in the plane of the sky. VLA observations by \citet{2006AJ....131..939Z} show multiple sources in this region. The spectral energy distribution of the source associated with the outflow observed by \citet{2004ApJ...616L..23B} is modelled by  a cm-wave component with a flat or slowly rising spectrum, which may be a thermal jet or stellar wind, while the mm/submm component rises rapidly with frequency and is probably thermal emission from dust.

\citet{2007A&A...466..215L} detected emission from methanol lines in the range 25 -- 255 GHz, with the characteristic intensity and linewidths of a hot core.  They fitted the observed lines with a three-component model, representing bulk emission from the molecular clump with T$_K$ = 21.8K, n(H$_2$) = $2.8 \times 10^6$  cm$^{-3}$ and v$_{lsr}$ = 32.9 km s$^{-1}$; an extended component with T$_K$ = 45K, n(H$_2$) = $7.0 \times 10^7$ cm$^{-3}$ and v$_{lsr}$ = 32.7 km s$^{-1}$; and a compact core with  T$_K$ = 300K and n(H$_2$) = $1.0 \times 10^8$ cm$^{-3}$ and v$_{lsr}$ = 32.7 km s$^{-1}$ .

In this paper we present the results of nine years of monitoring G12.89+0.49 at 6.7 and 12.2 GHz. The rate of observing was increased in September 2005 to better characterise the variability. We present detailed time series analysis to characterise the variability.

\section{Observations}

All observations were made with the Hartebeesthoek Radio Astronomy Observatory 26m telescope. The observing parameters of the monitoring program are given in Table~\ref{tab:instr}. Prior to 2003 only left circular polarisation was recorded. Observing took place at 1--2 week intervals.  The antenna surface at this time consisted of perforated aluminium panels.  The surface was  upgraded to solid panels during 2003-2004, resulting in increased efficiency at higher frequencies. However, the telescope pointing was affected until the telescope was rebalanced in April 2005 and a new pointing model was derived and implemented in September 2005. Pointing checks during observations were done by observing offset by half a beam-width at the cardinal points and fitting a 2-dimensional Gaussian beam model to the observed peak intensities.  The sources were observed with hour angle less than 2.3 hours to minimise pointing errors. Pointing corrections after the implementation of the new pointing model were typically around 4--10 \% at 12.2 GHz and 1--5 \% at 6.7 GHz. Any observations with pointing corrections greater than 30\% were excluded from the final dataset since it was found that these were invariably outliers in the time series. 

Amplitude calibration was based on monitoring of 3C123, 3C218 and Virgo A (which is bright but partly resolved), using the flux scale of \citet{Ott94}.  The antenna temperature from these continuum calibrators was measured by drift scans.  Pointing errors in the north-south direction were measured via drift scans at the beam half power points and the on-source amplitude was corrected using the Gaussian beam model.  

The bright methanol maser source G351.42+0.64, which previous monitoring has shown to exhibit only small variations in the strongest features \citep{2004MNRAS.355..553G}, was observed in the same way as G12.89+0.49 to provide a consistency check on the spectroscopy.  These sources were observed frequently, usually on the same days as G12.89+0.49. The observations were filtered on system temperature to remove data affected by rainstorms. These were apparent as outliers in the time series of the system temperature and flux density. An instability in the 4.5 cm system temperatures, at a 5\% level, was seen in both the comparison and target sources between MJD 53620 and 53780, along with a moderately elevated system temperature.  It manifests in the time-series as a rapid, small amplitude variation with a time-scale of $\sim$ 6 hours. This behaviour ceased after the receiver went through a warm-up and cool-down cycle.

During 2003 the receivers were upgraded to dual polarisation and  observations were switched to a new control system and a new spectrometer.  Frequency-switching mode was used for all observations. Daily observations (whenever possible) were undertaken during September 2005 to September 2006.

\begin{table}
\begin{center}
\caption{Observing parameters. Average system temperatures and rms noise are given.}
\label{tab:instr}
\begin{tabular}{lrrrrr}
\hline
Observation dates	&	bw	&	chan.	& vel. res 		&	T$_{sys}$	&	rms	\\
			&	(MHz)	&		&	km s$^{-1}$	& (K)		&	(Jy)	\\
\hline
\multicolumn{6}{c}{Rest frequency: 6.668518 GHz}\\
1999/01/18 -- 2003/01/29	&	0.64	&	256	&0.112	&51	&0.5		\\
2003/08/24 -- 2007/11/30	&	1.00	&	1024	&0.044	&70	&0.4		\\
\\
\multicolumn{6}{c}{Rest frequency: 12.178593 GHz}\\
2000/01/30 -- 2003/04/07	&	0.64	&	256	&0.062	&139	&2.0		\\
2003/08/12 -- 2007/12/01	&	1.00	&	1024	&0.048	&99	&0.3		\\
\hline
\end{tabular}
\end{center}
\end{table}

\section{Results and analysis}

\subsection{Comparison source G351.42+0.64}

\begin{figure}
\resizebox{\hsize}{!}{\includegraphics[clip,angle=0]{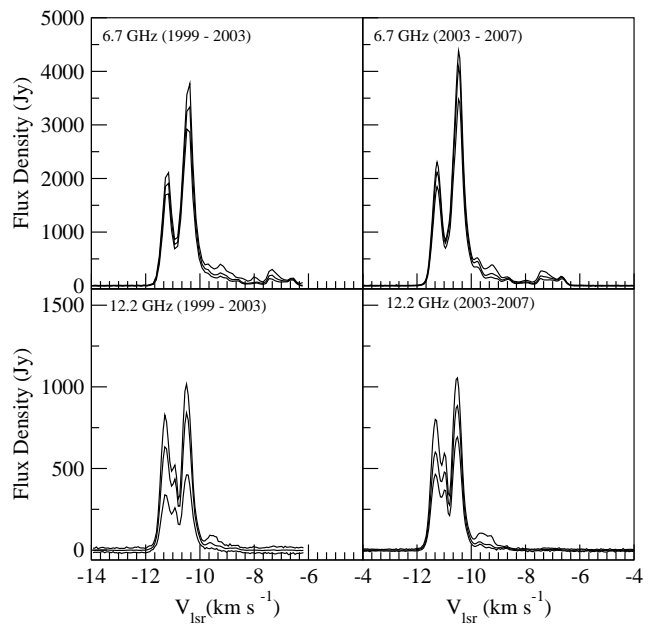}}
\caption{Range of variation of the masers in G351.42+0.64. In each panel, the upper curve shows the upper envelope, the middle curve is the averaged spectrum and the lower curve shows the lower envelope. } 
\label{fig:g3514-env} 
\end{figure}

\begin{figure*}
\resizebox{\hsize}{!}{\includegraphics[clip,angle=270]{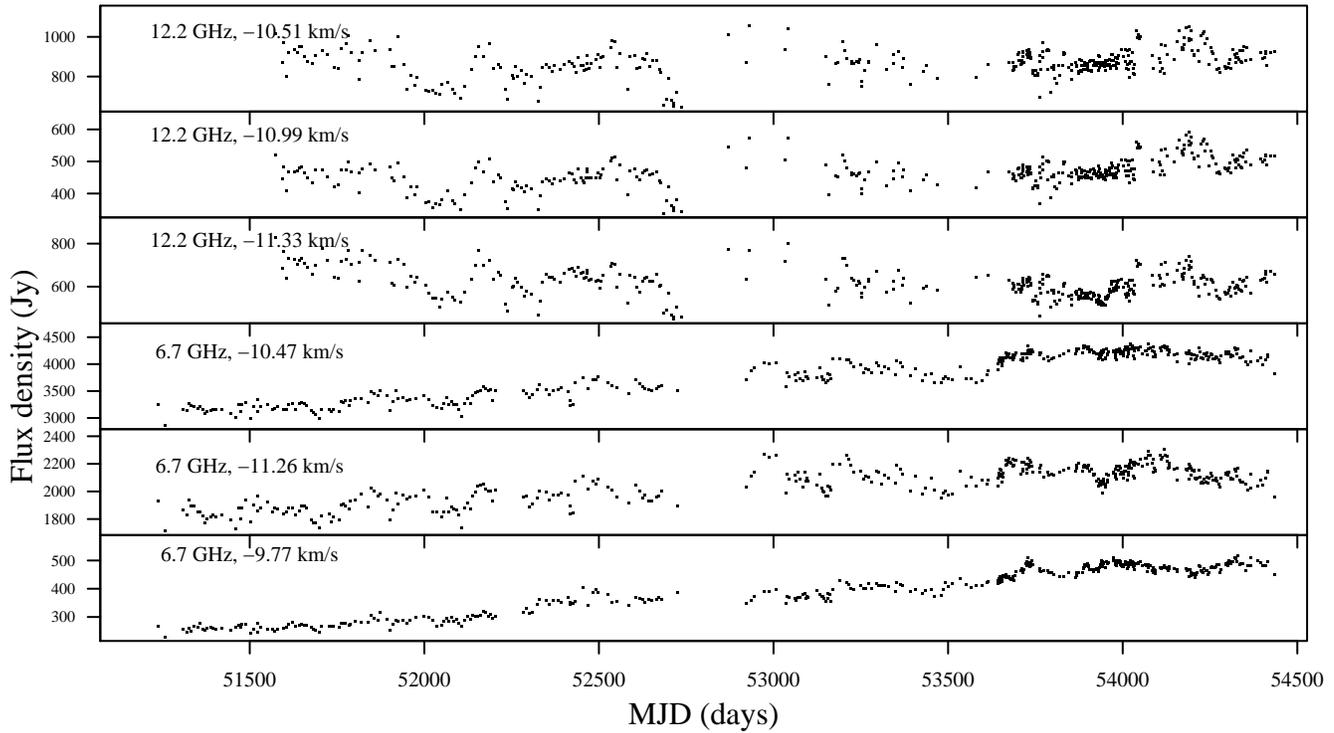}}
\caption{Time series for selected features in G351.42+0.64.} 
\label{fig:g3514-ts} 
\end{figure*}

\begin{figure*}
\resizebox{\hsize}{!}{\includegraphics[clip,angle=270]{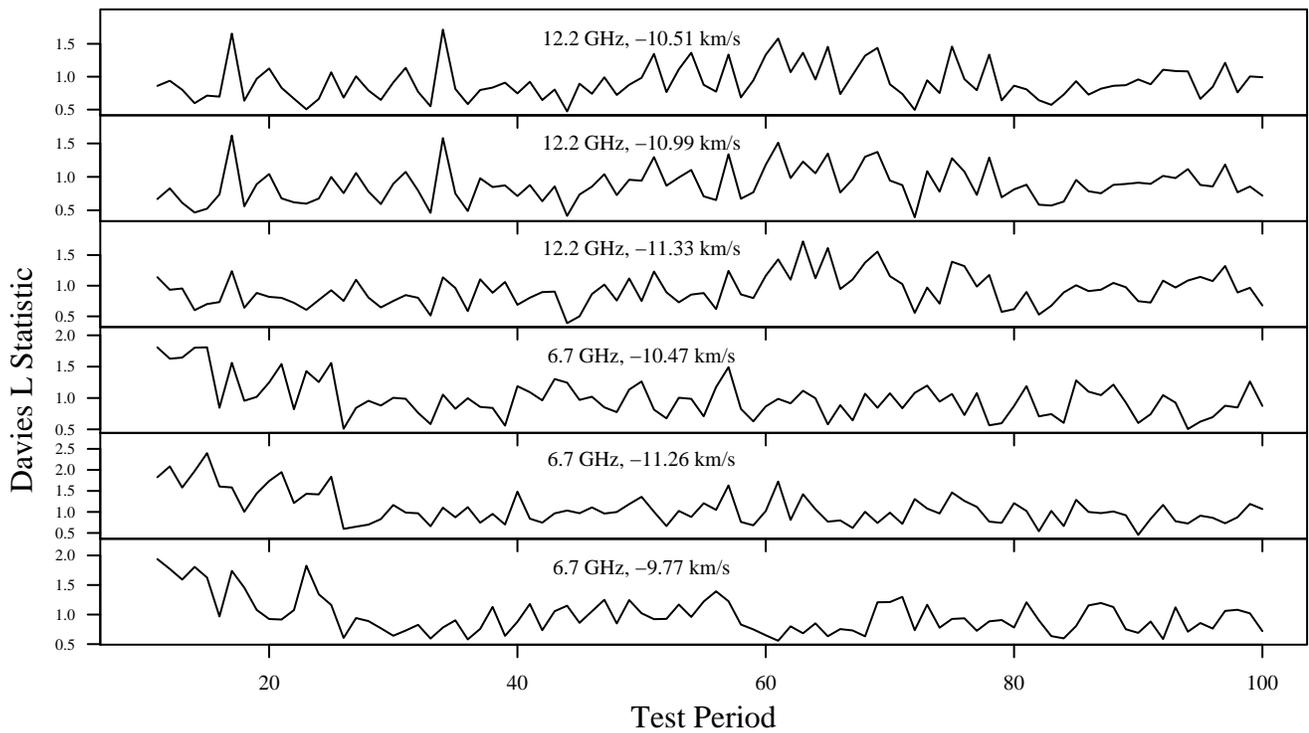}}
\caption{Result of epoch-fold period search on G351.42+0.64.} 
\label{fig:g3514-ef}
\end{figure*}

\begin{figure*}
\resizebox{\hsize}{!}{\includegraphics[clip,angle=270]{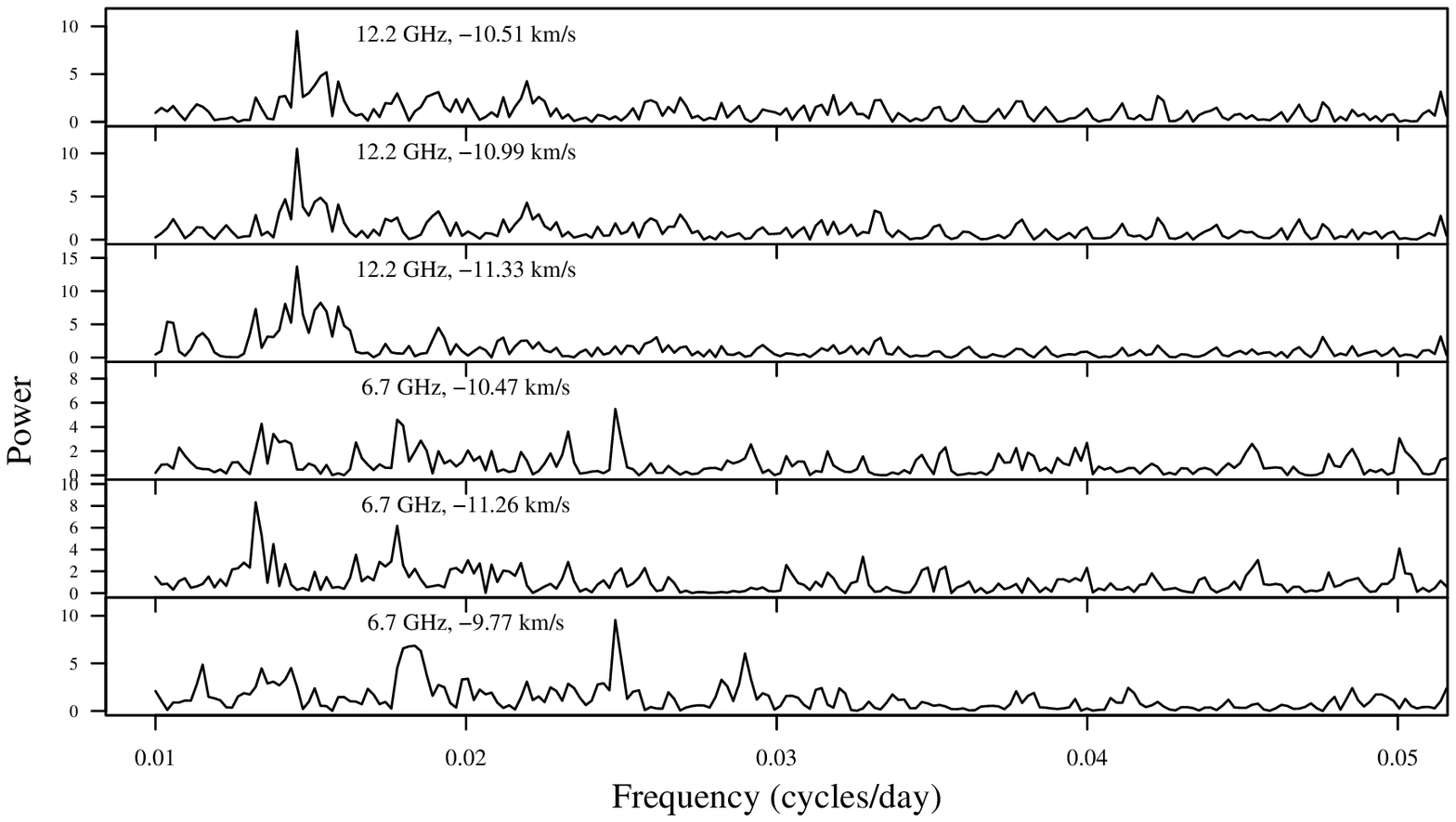}}
\caption{Scargle periodogram for G351.42+0.64. The 3$\sigma$ Z$_0$ level is 12.82.} 
\label{fig:g3514-dft}
\end{figure*}

G351.42+0.64 was chosen as a comparison source because its high intensity enables it to be observed using short integration times, and it was known to be only moderately variable.  Figure~\ref{fig:g3514-env} shows the range of variation in all spectral channels at 6.7 and 12.2 GHz for this source.  The upper envelope is calculated by finding the maximum value in the time series in each spectral channel, likewise the lower envelope consists of the minimum value recorded in each channel. Visual inspection  of the  upper and lower envelopes shows that all of the features at both transitions have shown significant variability during the monitoring period.  Figure~\ref{fig:g3514-ts} shows the time series in a few of the velocity channels at 6.7 and 12.2 GHz. Moderate, irregular variability is observed in all of the channels. The 6.7 GHz maser at -9.77 km s$^{-1}$  has been gradually increasing in intensity since the start of the monitoring program and its variations appear to be uncorrelated with the behaviour seen in other channels.

A period search was done on this source as a means of verifying that any periodic signal found in the target source is unique to the source and not an artifact of the telescope or its sub-systems. Two methods of period search were used: epoch-folding using the test statistic $L$ of \citet{Dav90} and the discrete Fourier transform (DFT) \citep{Sca82}. The time series were detrended using a first-order polynomial before carrying out the period search in order to remove any long-term trends. Figure~\ref{fig:g3514-ef} shows the results of the epochfold in the period range of interest, on selected spectral channels, and Figure~\ref{fig:g3514-dft} shows the discrete Fourier transform in the same frequency range. There is no signal exceeding the 3-$\sigma$ level of 12.82 but there does appear to be some excess power at 2-$\sigma$ at a frequency of 0.0156 cycles/day (period = 64 days) in the 12 GHz power spectra.  This may well be spurious since the epoch-fold search shows no evidence of a signal at 64 days. 

\subsection{G12.89+0.49}

\begin{figure}
\resizebox{\hsize}{!}{\includegraphics[clip,angle=0]{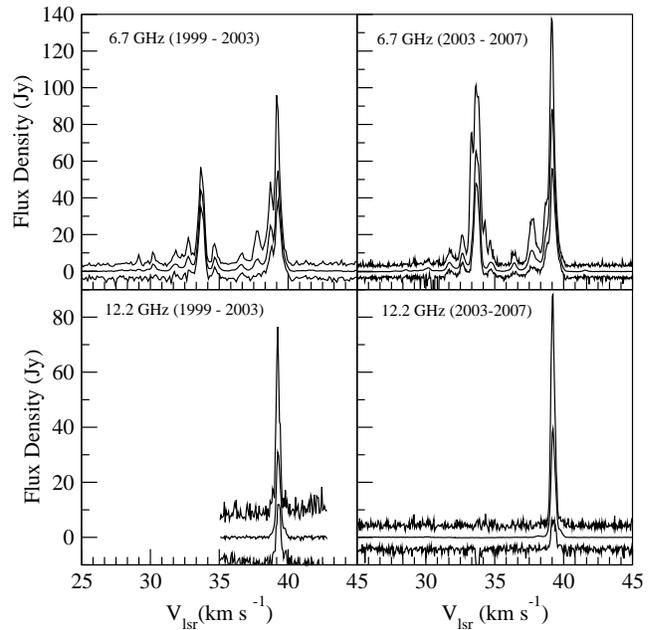}}
\caption{Range of variation of the masers in G12.89+0.49.  In each panel, the upper curve shows the upper envelope, the middle curve is the averaged spectrum and the lower curve shows the lower envelope.} 
\label{fig:g0128-env} 
\end{figure}

\begin{figure*}
\resizebox{\hsize}{!}{\includegraphics[clip,angle=270]{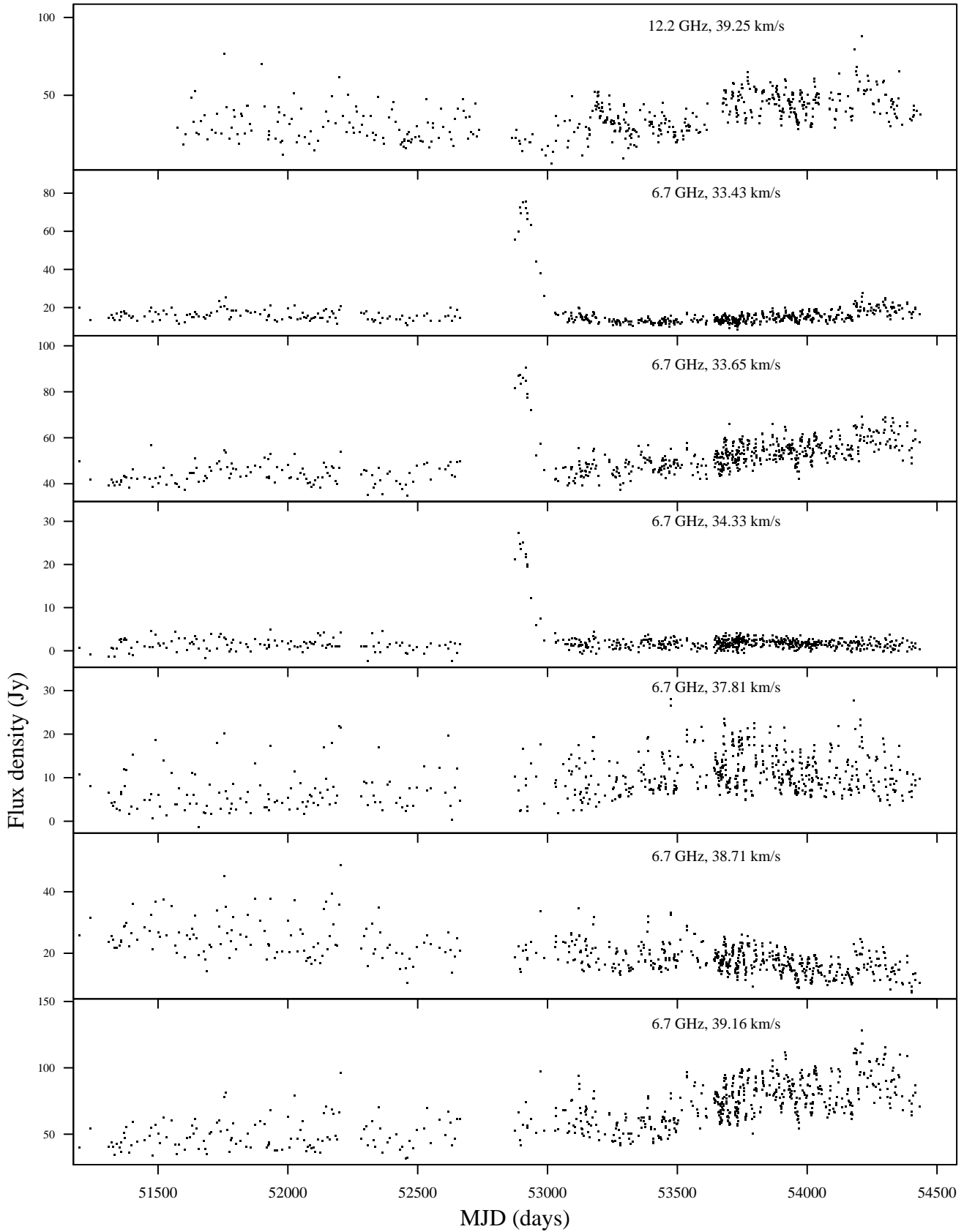}}
\caption{Time series for selected features in G12.89+0.49. Note that the top panel shows the 12.2 GHz feature while all the others are at 6.7 GHz.} 
\label{fig:g0128-ts-full} 
\end{figure*}

\begin{figure*}
\resizebox{\hsize}{!}{\includegraphics[clip,angle=270]{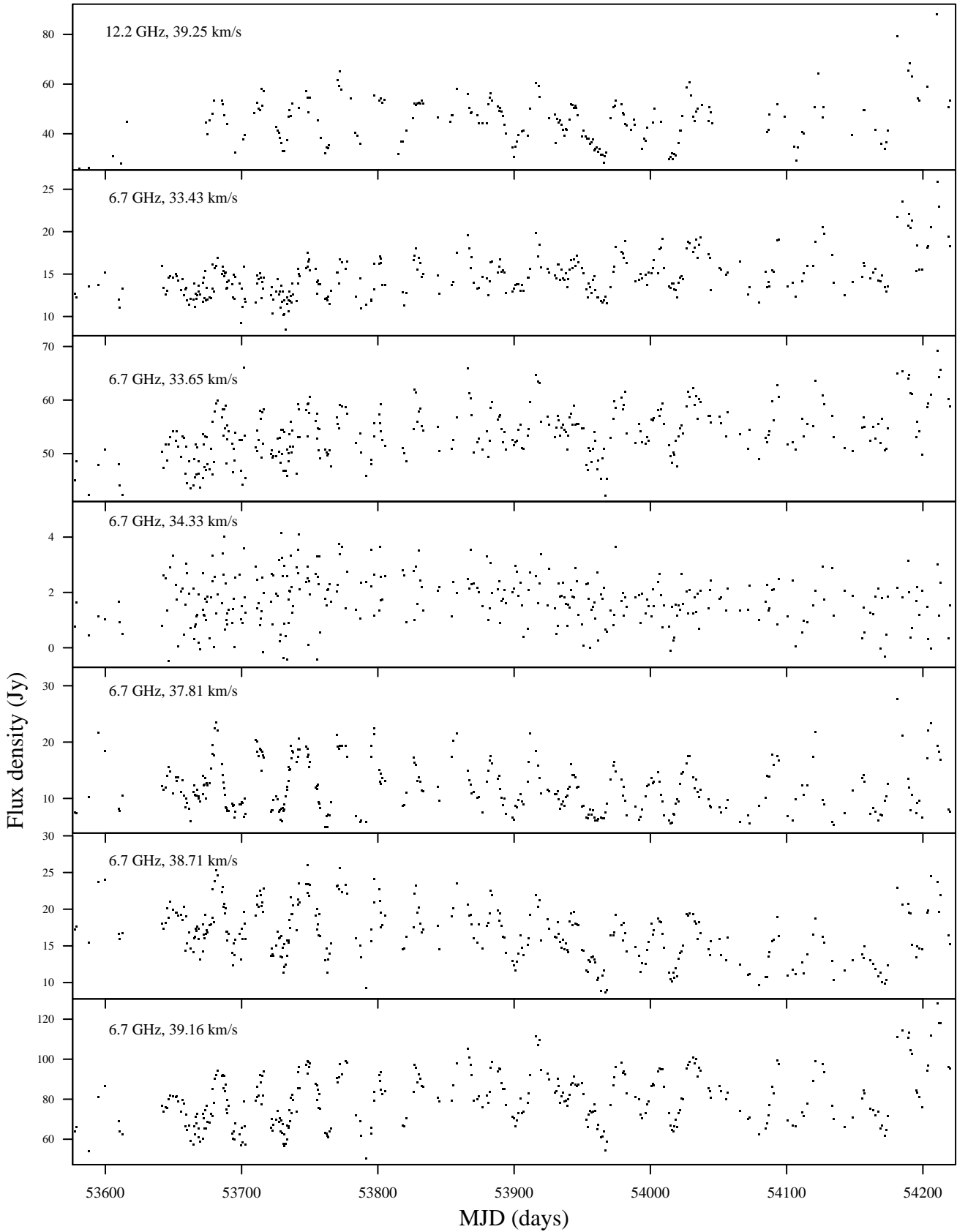}}
\caption{Time series for selected features in G12.89+0.49 covering the period over which daily monitoring was done.} 
\label{fig:g0128-ts-zoom} 
\end{figure*}

The range of variation for all spectral channels in G12.89+0.49 at 6.7 and 12.2 GHz is shown in Figure~\ref{fig:g0128-env}. Peak channels in the spectrum were extracted for time series analysis and additional features, which may be hidden in the line wings of the dominant peaks, were identified by comparing the upper envelope to the average spectrum to find additional flaring features. These velocity channels show close correspondence to the individual maser features mapped by \citet{Wal98}. Figure~\ref{fig:g0128-ts-full} shows the time-series of the stronger features. \textit{Plots of the time series of all of the identified features will be available online.} Only one spectral feature is seen at 12.2 GHz. The 6.7 GHz spectrum is comparatively rich.  Of note is the change in the spectrum at 6.7 GHz in the nine years of observations. The peak at 30.18 km s$^{-1}$ has decreased in intensity. A range of  features at $\sim$ 33 -- 35 km s$^{-1}$ show the largest range of variation.  Examination of the time series  shows that these features underwent a large flare in 2004.  The feature at 33.65 km s$^{-1}$ has been increasing in average intensity since the flare but the trend appears to have levelled off during 2007. The time series prior to September 2005  show a significant level of variability, but it is not until the start of daily monitoring that the nature of the variability becomes clear.  Rapid fluctuations can be seen in the closely-sampled segment (Figure~\ref{fig:g0128-ts-zoom}, \textit{more features shown in the online version of the figure}). It is clear that the source exhibits rapid, regular  flaring in most, if not all, of the spectral features. 

\begin{figure*}
\resizebox{\hsize}{!}{\includegraphics[clip,angle=270]{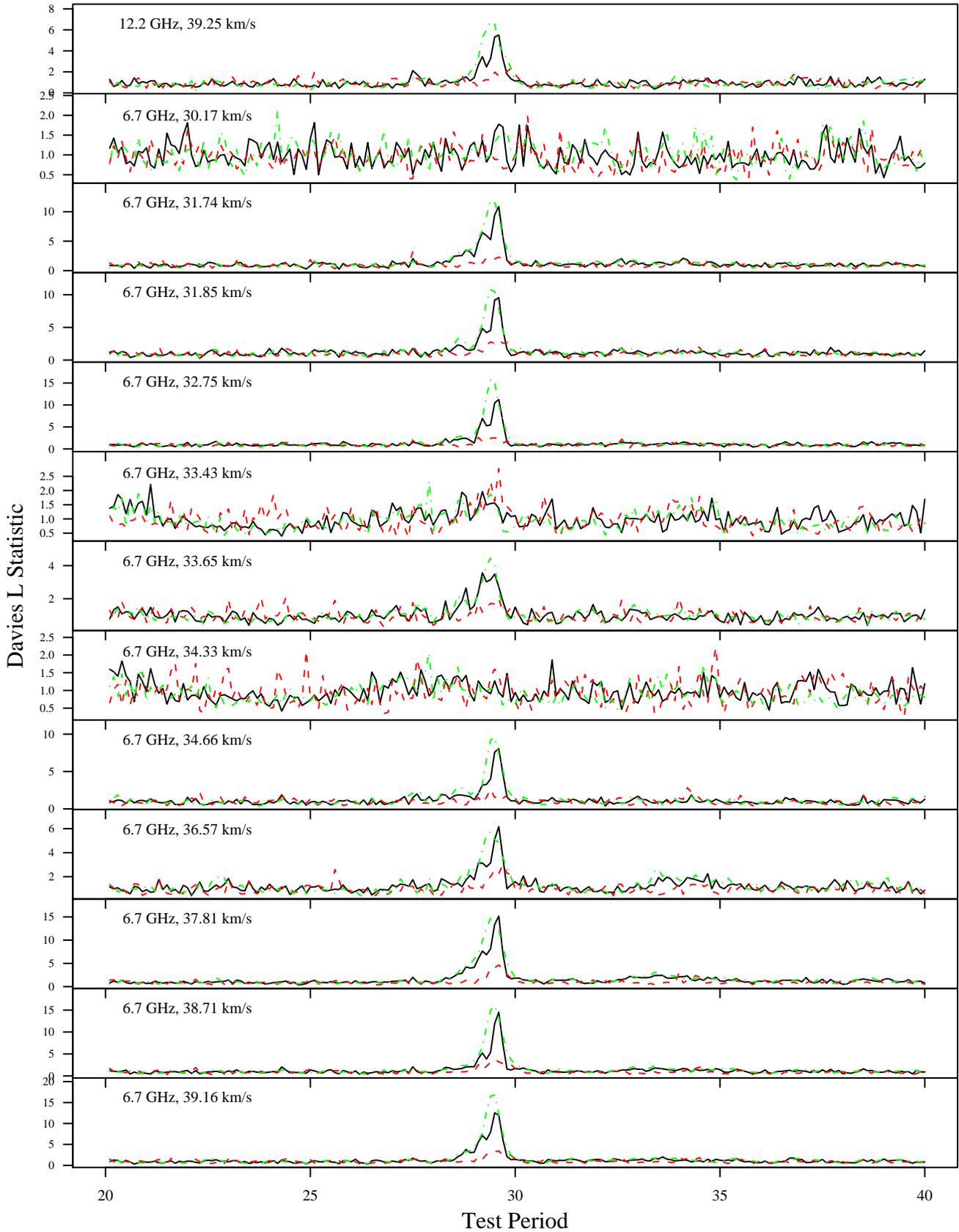}}
\caption{Comparison of epoch-fold search over different time spans. The solid line shows the results for the full time span, the dashed line shows the range 1999-2003 (which was undersampled), the dash-dot line shows the range 2003-2007.} 
\label{fig:ef-comp}
\end{figure*}

\begin{figure*}
\resizebox{\hsize}{!}{\includegraphics[clip,angle=270]{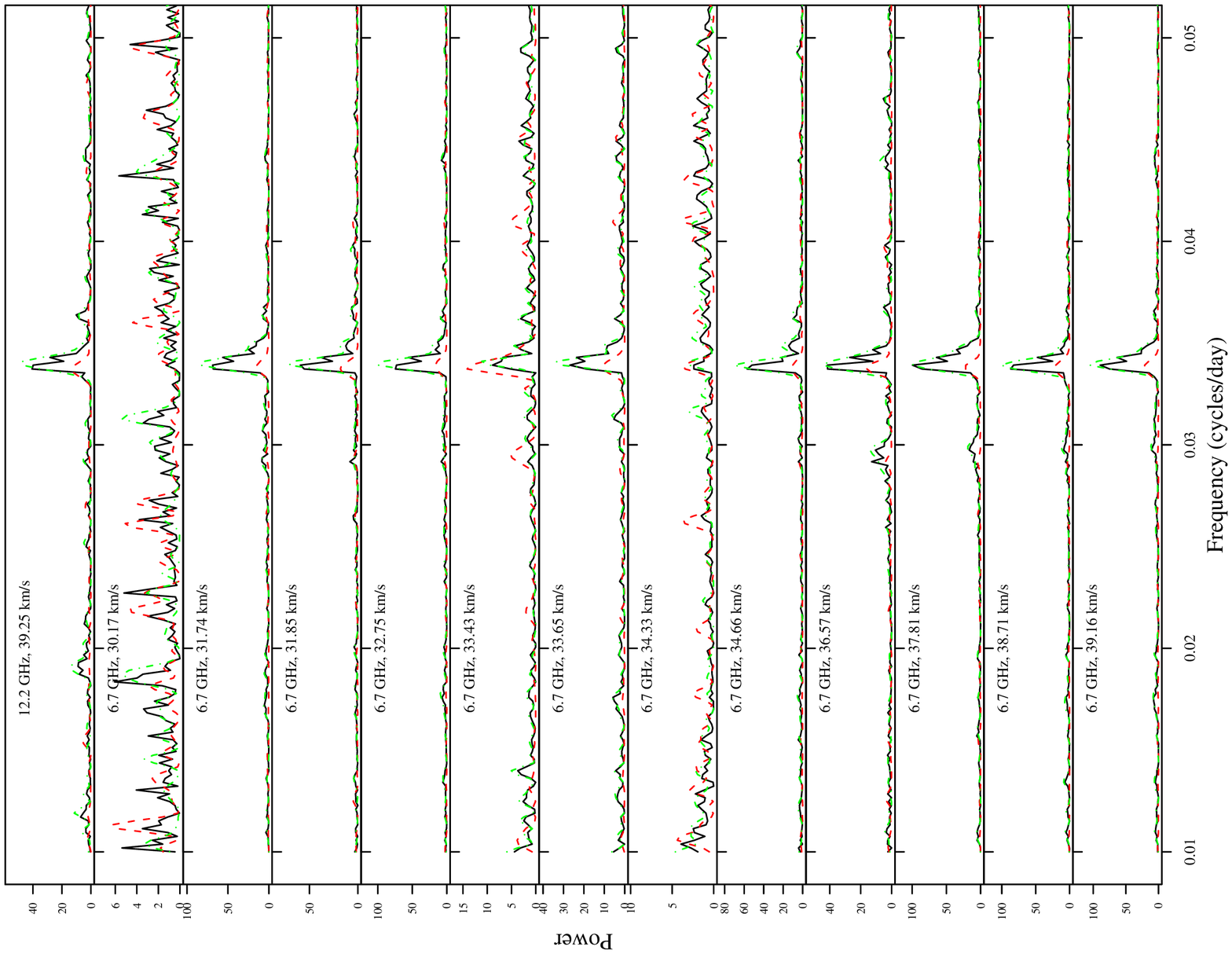}}
\caption{Comparison of Scargle periodograms over different time spans. The solid black line shows the results for the full time span, the dashed red line shows the range 1999-2003 (which was undersampled), the green dash-dot line shows the range 2003-2007. The 3$\sigma$ Z$_0$ level is 12.82.} 
\label{fig:g0128-dft}
\end{figure*}

\subsubsection{Periodicity and flaring}

The results of period searches using the Davies \textit{L}-statistic and DFT are shown in Figures~\ref{fig:ef-comp} and \ref{fig:g0128-dft} respectively. The searches were done over the full time span, as well as separately on the 1999 -- 2003 and 2003 -- 2007 segments, which have different sampling rates. The same signal appears to be present in the under-sampled time series as in the well-sampled segment.  A summary of peak periods found over various time segments and using the two methods, as well as representative S/N, are given in Table~\ref{tab:stats}.  The periods found range from 29.29 to 29.78 days with a bias towards a slightly longer period in the earlier time segment. However, since the earlier data are undersampled and noisier, it is not clear whether this is significant.  The weighted mean (using the S/N of the time series to give the weights) for the epoch-folding period search is 29.51 days with a standard deviation of 0.09 days.  The weighted mean for the DFT-based period search is 29.53 with a standard deviation of 0.10 days.  The most likely period is thus derived to be 29.5 $\pm$ 0.1 days. 

\begin{figure}
\resizebox{\hsize}{!}{\includegraphics[clip,angle=270]{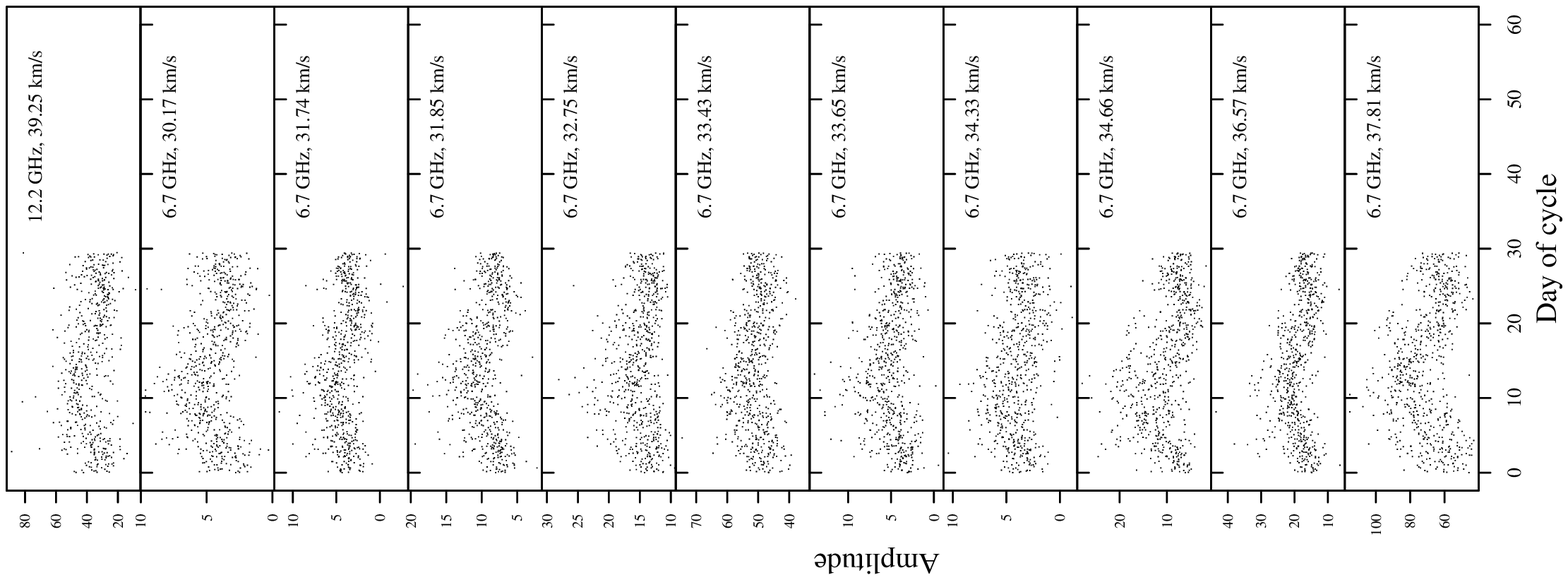}}
\caption{time series folded modulo 29.5 days.} 
\label{fig:g0128-fold}
\end{figure}

\begin{figure}
\resizebox{\hsize}{!}{\includegraphics[clip,angle=270]{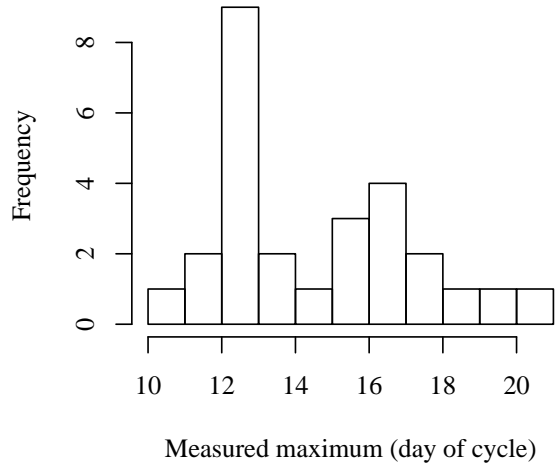}}
\caption{Histogram of measured peak flare times for 27 well-sampled cycles.} 
\label{fig:max-hist}
\end{figure}

\begin{figure}
\resizebox{\hsize}{!}{\includegraphics[clip,angle=270]{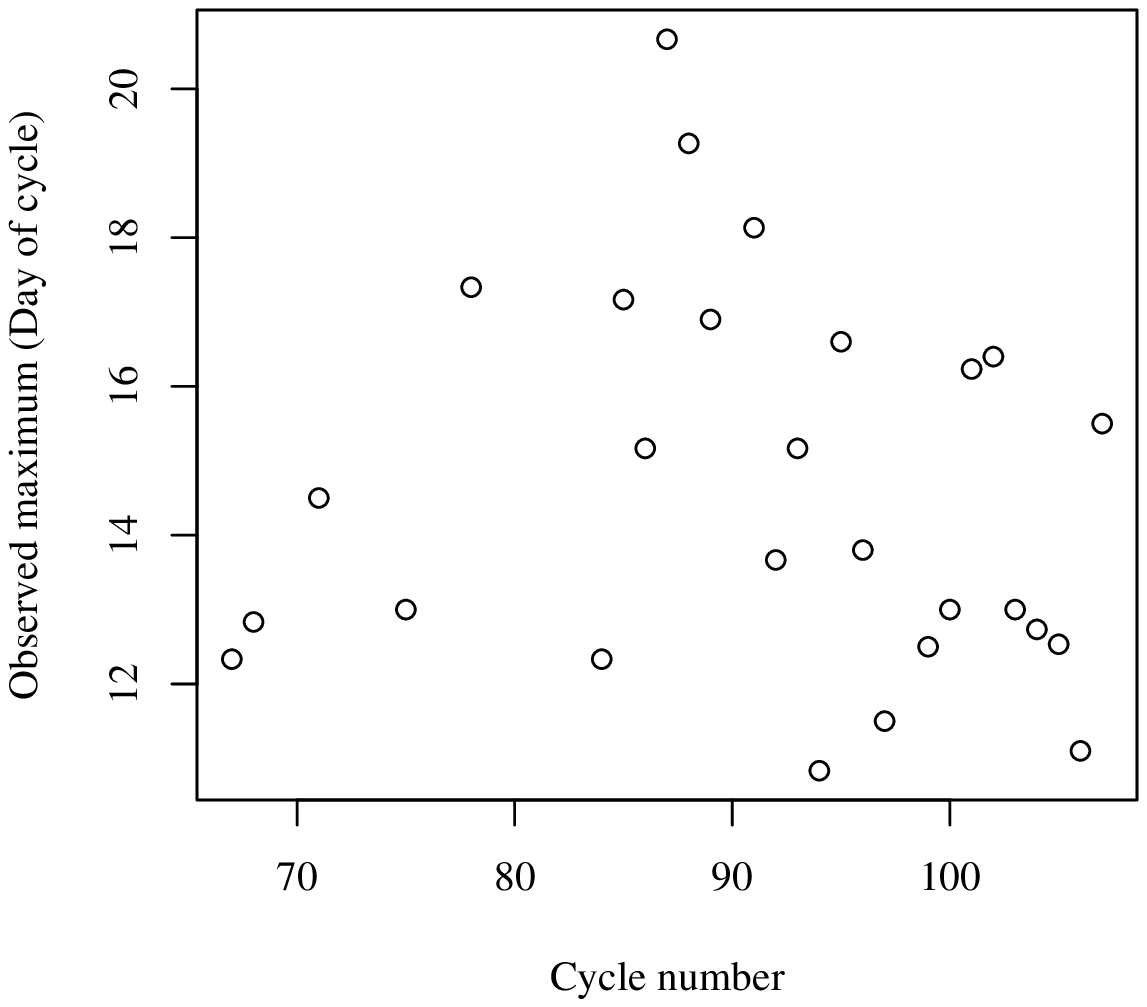}}
\caption{Measured peak flare time for 27 well-sampled cycles, as a function of cycle number.} 
\label{fig:cycle-max}
\end{figure}

Figure~\ref{fig:g0128-fold} shows the effect of folding the time series modulo 29.5 days. The start of the cycle was chosen arbitrarily to roughly centre the flares in the plots. If the source were strictly periodic, the cycles would overlay each other to within the noise of the time series, showing the characteristic pulse shape.  It is clear that the cycles do not repeat exactly, undergoing flares of different amplitudes and in the case of at least one cycle, not showing a flare at all. It is interesting to note that the minima of the cycles do repeat very well. The time of maxima of the flares in each cycle  (assuming a period of 29.5 days) for the 39.16 km s$^{-1}$ 6.7 GHz maser were fitted to characterise the spread in flare peaks.  This was done for those cycles where sufficient points were observed to enable a reasonable fit using a second-order polynomial over the duration of the flare. It was possible to  measure the peak of the flare for 27 cycles in the data taken after August 2003. In one case (cycle 74) there did not appear to be a flare despite frequent observations throughout the cycle. The flare is present but not very strong or well defined in cycle 95. A histogram of the results is shown in Figure~\ref{fig:max-hist}. The distribution does not appear to be Gaussian but there are too few cycles measured to make a definitive statement. There is a clustering at $\sim$ 12.5 days but there are a significant number of flares which peak later into the cycle. Figure~\ref{fig:cycle-max} shows the measured time of maximum as a function of cycle number i.e the progression of the flares with time.  The occurence of the peak of the flare appears to be random with no obvious trends or systematic drifts. There does not appear to be any correlation of the flare amplitude and the phase. The flares always peak within an interval that is approximately 1/3 of the period, hence the well defined minima in the folded time series.

The typical rise times of the flares can be used to estimate the linear scale of the masers.  The 39.16 km s$^{-1}$ 6.7 GHz feature takes roughly 12 days on average from the minimum point in its cycle to reach its maximum. The maser region undergoing the flare therefore cannot have a linear extent more than 12 light days or 2 074 AU.   Assuming a distance of 3.6 kpc and an average increase in flux density of 30 Jy, the lower limit on the brightness temperature is  $\sim 4 \times 10^{11}$ K. This is consistent with the highest brightness temperatures that can be reproduced by the model of \citet{2005MNRAS.360..533C}. VLBI observations typically reveal maser spot sizes of the order of a few hundred AU \citep[e.g.][]{Min02} so it is possible that the brightness temperature of the maser could be even higher.

\afterpage{\clearpage}

\subsubsection{Spectral variations}

\begin{figure*}
\resizebox{\hsize}{!}{\includegraphics[clip,angle=0]{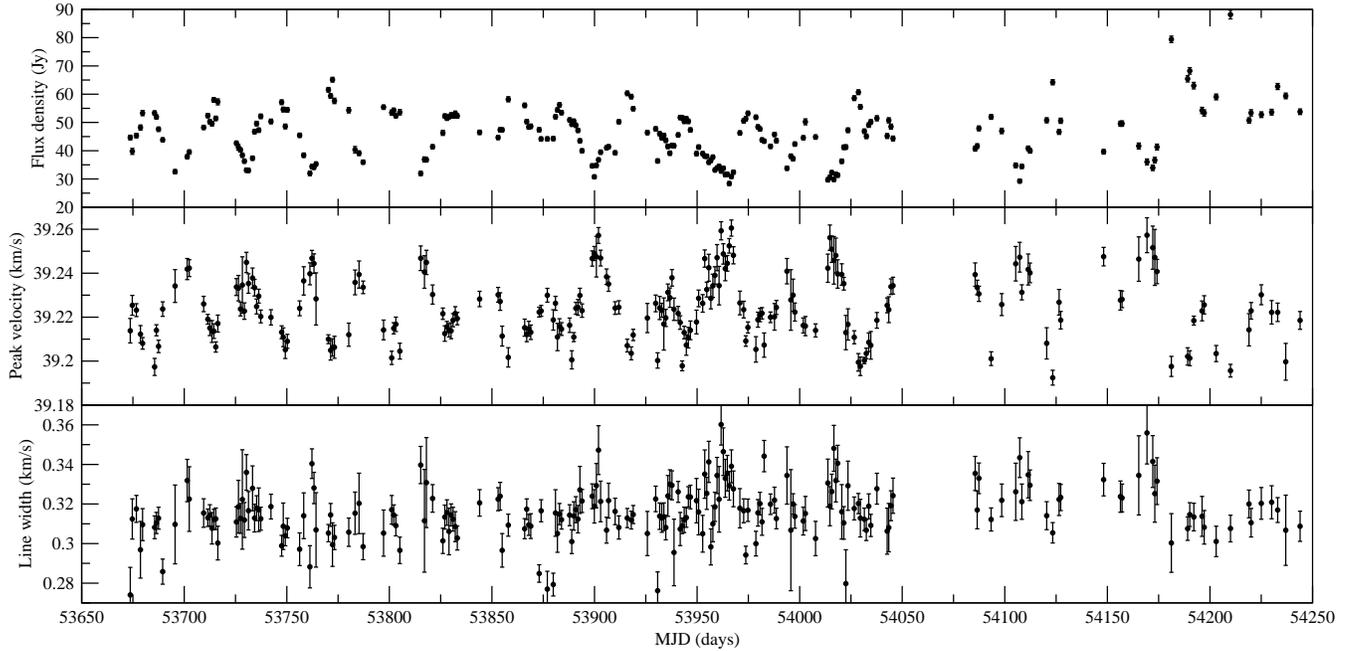}}
\caption{Results of Gaussian fits to the dominant spectral peak at 12.2 GHz.  The top panel shows the flux density of the spectral channel at 39.23 km s$^{-1}$, the middle and bottom panels shows the change in the Gaussian peak position and width, respectively, of the spectral feature.} 
\label{fig:vel-shift-12}
\end{figure*}

\begin{figure*}
\resizebox{\hsize}{!}{\includegraphics[clip,angle=0]{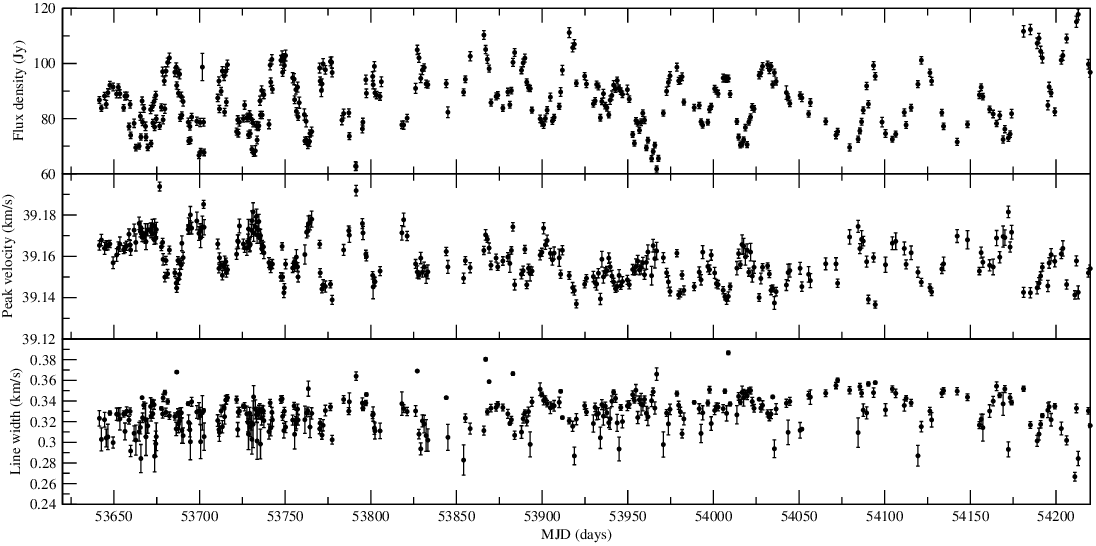}}
\caption{Results of Gaussian fits to the dominant spectral peak at 6.7 GHz.  The top panel shows the flux density of the spectral channel at 39.16 km s$^{-1}$, the middle and bottom panels shows the change in the Gaussian peak position and width, respectively, of the spectral feature.} 
\label{fig:vel-shift-6}
\end{figure*}

\begin{figure*}
\resizebox{\hsize}{!}{\includegraphics[clip,angle=0]{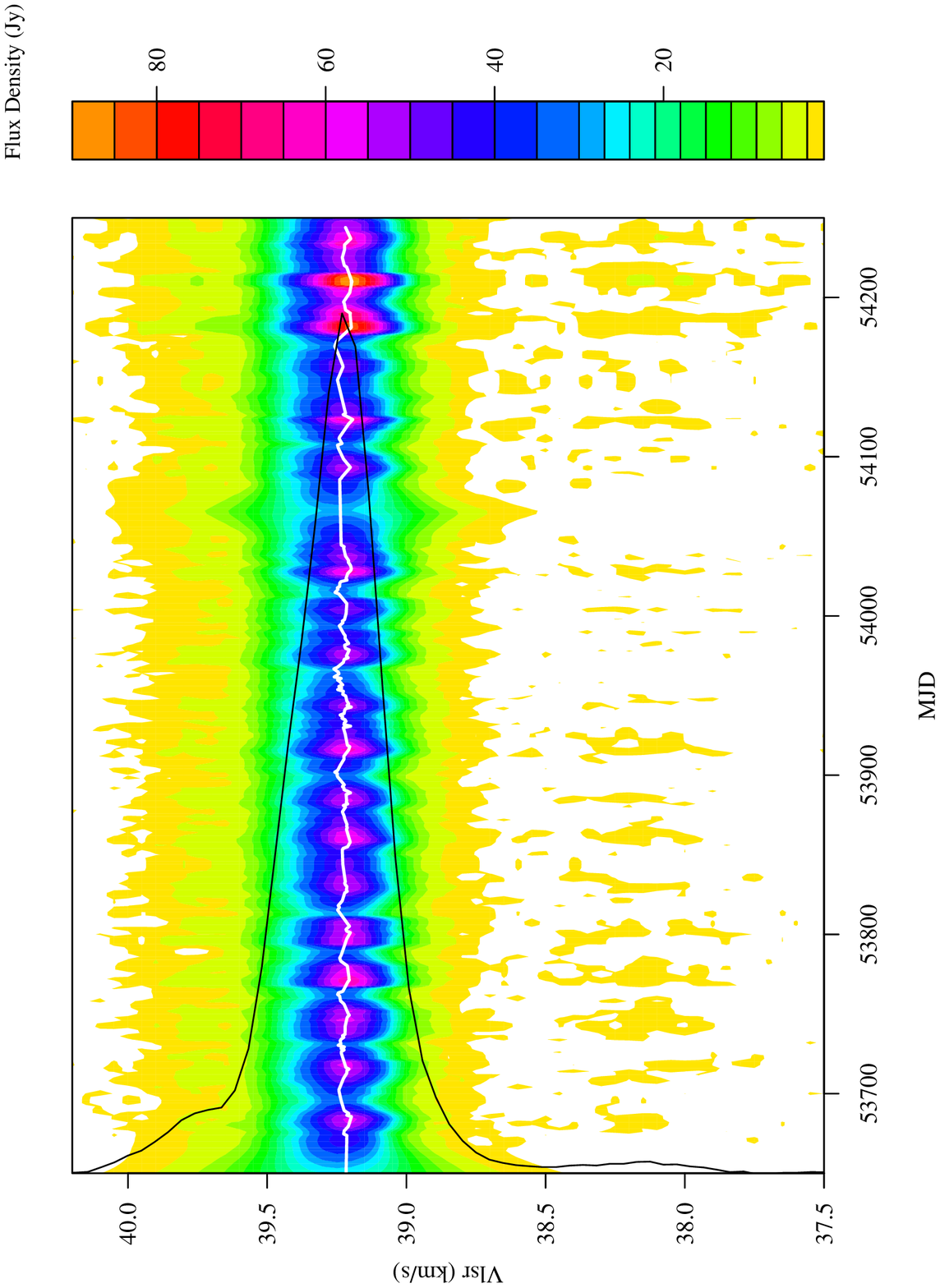}}
\caption{Contour plot of flux density over velocity and time for the 12.2 GHz transition over the full velocity range of emission.  The black curve is a scaled average spectrum. The white curve shows the result of fits to the position of the dominant Gaussian in the spectrum.} 
\label{fig:contour-ts-122}
\end{figure*}

\begin{figure*}
\resizebox{\hsize}{!}{\includegraphics[clip,angle=0]{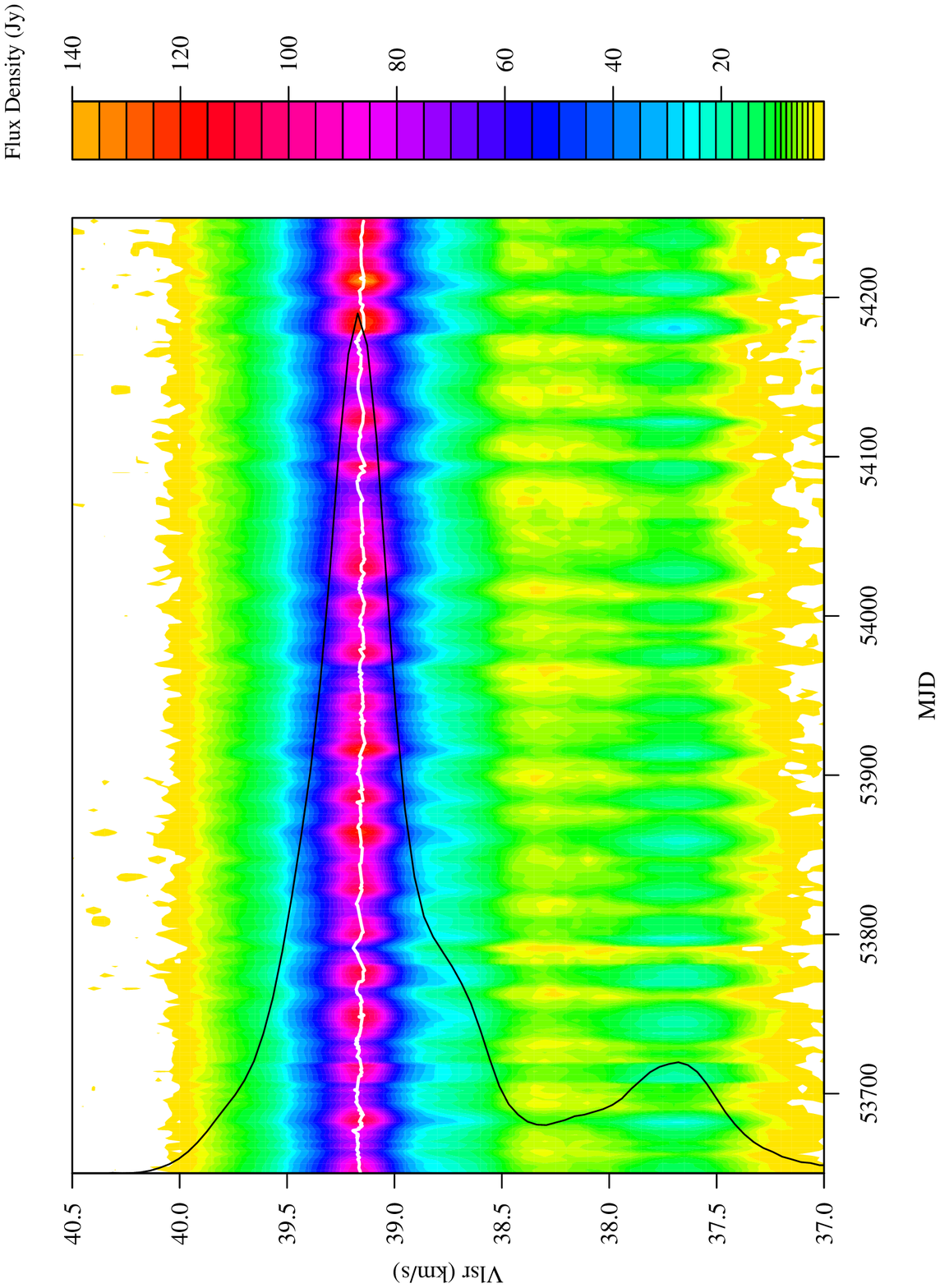}}
\caption{Contour plot of flux density over velocity and time for the 6.7 GHz transition for the velocity range 37 to 40.5 km s$^{-1}$.  The black curve is a scaled average spectrum. The white curve shows the result of fits to the position of the dominant Gaussian in the spectrum.} 
\label{fig:contour-ts-67a}
\end{figure*}

\begin{figure*}
\resizebox{\hsize}{!}{\includegraphics[clip,angle=0]{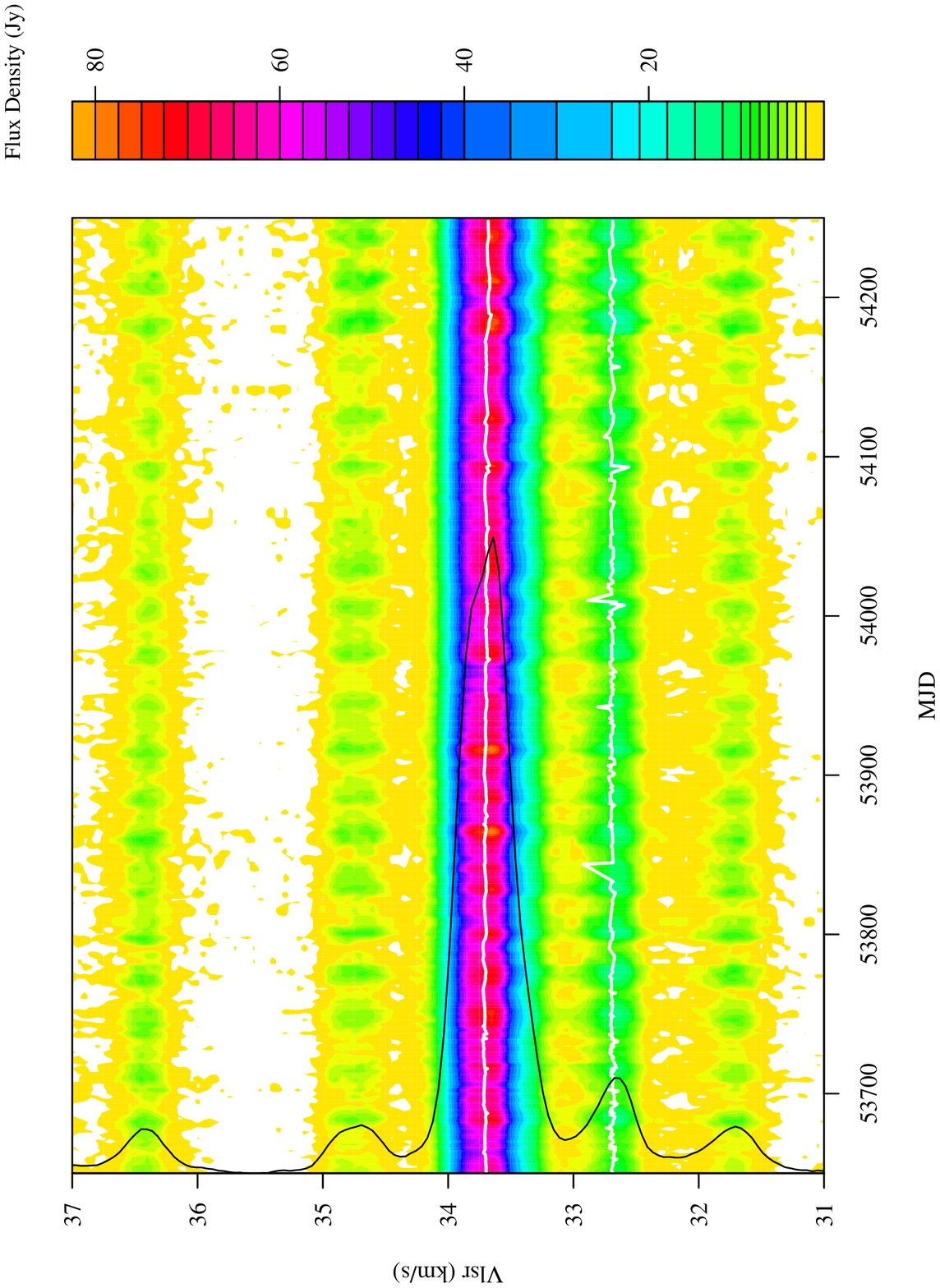}}
\caption{Contour plot of flux density over velocity and time for the 6.7 GHz transition for the velocity range 31 to 37 km s$^{-1}$.  The black curve is a scaled average spectrum. The white curves shows the result of fits to the position of the dominant Gaussians in the spectrum.} 
\label{fig:contour-ts-67b}
\end{figure*}

Gaussian profiles were fitted to each spectrum to check whether any shifts in velocity occurred.  The 12.2 GHz spectrum was fitted using three Gaussian lines.  The two weaker lines are only detected during flares and the S/N of the fit is poor.  The 6.7 GHz spectrum was fitted using thirteen Gaussian lines. The quality of the fit is good on the two dominant peaks but the weaker peaks are very broad and not well described by Gaussian profiles.  Figure~\ref{fig:vel-shift-12} shows the position and width of the Gaussian line fit over time compared to the flux density in the 39.23 km s$^{-1}$ spectral channel at 12.2 GHz and Figure~\ref{fig:vel-shift-6} shows the same for the equivalent 6.7 GHz peak. Shifts of less than one channel width in the Gaussian position are clearly seen and strongly correlated with the amplitude variation of the maser peak. There are small changes in the line-widths as well. In the light of this result, it is necessary to look at the variations in the entire spectral profile and not just a single channel.  Figures~\ref{fig:contour-ts-122}, \ref{fig:contour-ts-67a} and \ref{fig:contour-ts-67b} show the intensity changes in all of the spectral channels during the daily monitoring campaign. The 6.7 GHz spectrum data has been split into two velocity ranges in order to view finer details. The Gaussian position fit has been added to the plot to verify the validity of the result. The shift in the profile is clearly visible with the maximum moving towards the red-shifted side of the peak at ~39.2 km s$^{-1}$ for both transitions. A similar but smaller shift is seen in the 34.65 km s$^{-1}$ feature.  This shift may be a real change in the velocity of the maser feature or it could be the result of a superposition of masers with similar velocities but varying flare amplitudes. Strong correlations between spectral variability and intensity variations are indicative of spectral blending \citep{2007MNRAS.376..549M,2009MNRAS.392.1339M}. Cross-correlations of the fitted line velocity against the flux density in the peak channels give a correlation coefficient of 0.64 at 12.2 GHz (with a lag of two days between the maximum velocity shift and the maximum of the flare in the peak channel) and 0.86 at 6.7 GHz (with no lag). Thus the velocity shift is most probably caused by spectral blending. The correlation with the line-width change is poor, with a coefficient of 0.1 or less. This may be because the error on the line-width fit is much higher than for the peak velocity fit.

\subsubsection{Time delays}

\begin{figure}
\resizebox{\hsize}{!}{\includegraphics[clip,angle=270]{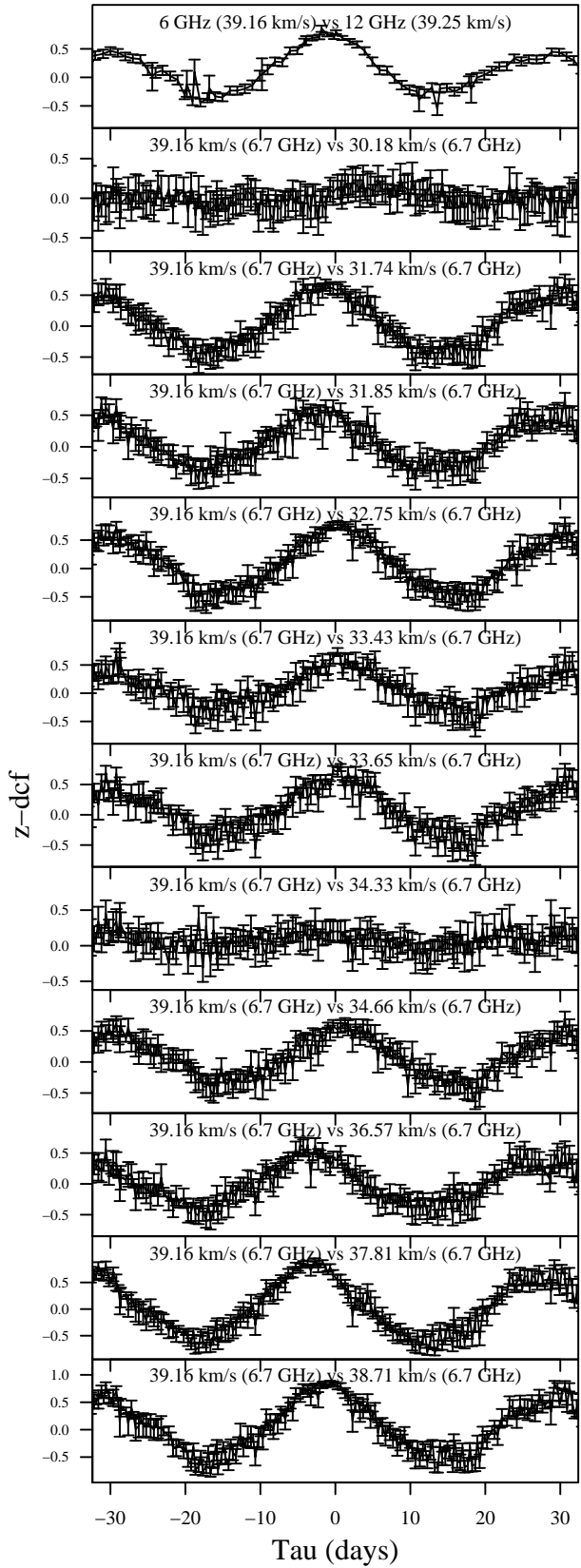}}
\caption{The z-transformed discrete correlation function for pairs of time series. The top panel is for the dominant peak (39 km s$^{-1}$) at 12.2 and 6.7 GHz. The subsequent plots are for cross-correlations at 6.7 GHz of the main peak against all other spectral features. The velocity channels are given in the legends.}
\label{fig:g0128-zdcf}
\end{figure}

Time-delays between features were checked for using the z-transformed\footnote{This is the Fisher z-transform.} discrete correlation function (z-dcf) code of \citet{Ale97}. There is a trade-off between increasing the S/N of the correlation by using more points per bin and the size of the bins used.  The greater the number of points per bin the better the S/N but the delay then cannot be determined accurately.  A minimum of 15 points per bin was used, giving a bin size of 0.83 days for the cross-correlation with the 12.2 GHz feature and 0.34 days for the 6.7 GHz features. It seems reasonable to assume an uncertainty in the delay of half the bin size for those features showing a high degree ($>$ 0.5) of correlation. The data for 2003 -- 2007 were used for this analysis because of the higher sampling rate. The results of cross-correlating with the main 6.7 GHz feature at 39.16 km s$^{-1}$ are shown in Figure~\ref{fig:g0128-zdcf}. The uncertainty in the cross-correlation is calculated by running 1000 Monte Carlo simulations for each pair of observed light curves. A sub-range of the full correlation function is shown for clarity but aliasing at multiples of the period are seen for all the peaks with a periodic signal, except for 30.18 km s$^{-1}$ and 34.33 km s$^{-1}$ which have poor S/N.  The detection of a correlation and the aliasing is a good indication that all the peaks have the same period as the main peak. The position of the maximum was found by fitting a third-order polynomial to the region around the peak. Table~\ref{tab:dcf} gives the fitted lag, the magnitude of the correlation and the uncertainty in the correlation from Monte Carlo simulation. A negative lag means that the maser feature listed in column one flares before the main 6.7 GHz peak.  The 12.2 GHz feature appears to flare 1.1 $\pm$ 0.4 days before the 6.7 GHz feature and there seem to be delays ranging between 0.3 to 3.7 $\pm$ 0.2 days between features at 6.7 GHz compared to the main peak. 

\begin{figure}
\resizebox{\hsize}{!}{\includegraphics[clip,angle=0]{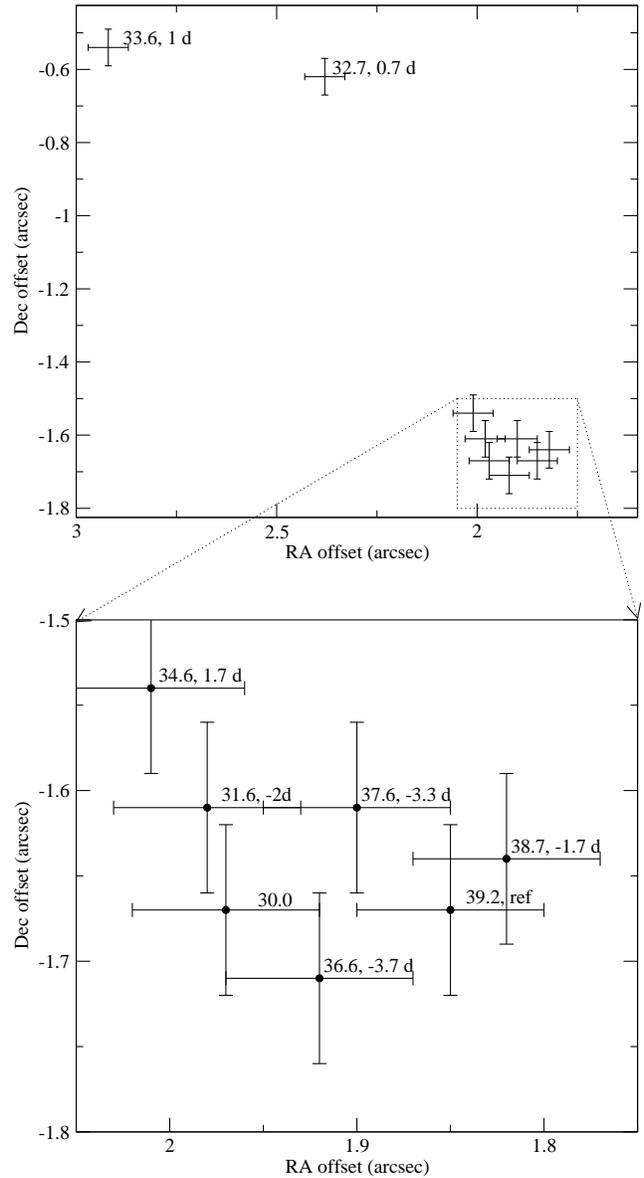}}
\caption{Position of maser spots as mapped by Walsh. The bottom panel is a close-up view of the cluster of spots shown in the top panel. The filled circles give the spot positions and the labels are the velocities and time delays relative to the reference feature at 39.2 km s$^{-1}$.} 
\label{fig:walsh} 
\end{figure}

It is necessary to take into account the maser spot distribution in order to attempt to understand these delays.  Figure~\ref{fig:walsh} shows the maser spot distribution as mapped by \citet{Wal98}. There does not appear to be any relation between the magnitude of the time delay and the position of the maser features in the plane of the sky. An attempt to  construct a 3-dimensional distribution based on the  projected maser separations and the time delays was unsuccessful unless a distance less than 200 pc was adopted. This distance is clearly too close to be plausible therefore the source of error is probably not in the commonly accepted distance of 3.6 kpc. The estimated positional errors of 0.05 arcsec in this spot map are significant compared to the separation between maser spots.  The estimated uncertainty of 0.2 days in the time-delays corresponds to $\sim$ 0.01 arcsec at a distance of 3.6 kpc, thus the positional errors dominate the calculation.

\begin{figure}
\resizebox{\hsize}{!}{\includegraphics[clip,angle=0]{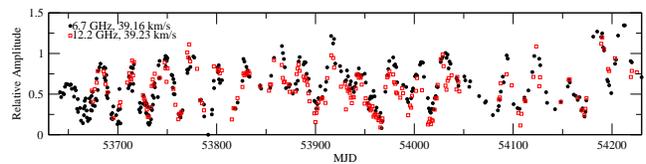}}
\caption{Comparison of flare behaviour at 12.2 and 6.7 GHz. Flare amplitudes have been normalised. }
\label{fig:jt}
\end{figure}

12.2 and 6.7 GHz methanol masers with the same  velocities have been found to be co-spatial down to milli-arsecond scales \citep{Men92, Min00} thus the clear delay of a day between flares in the 39.2 km s$^{-1}$ will be of significance to theoretical models of multiple transitions which  assume the same physical conditions for the maser species \citep[e.g.][]{Cra01,2001ApJ...554..173S}. The delay in the flare at 6.7 GHz implies that the bulk of the 12.2 GHz emission arises approximately 190 AU closer to the regularly varying photon source than the 6.7 GHz masers. Physical conditions may vary slightly betweeen the two transitions in this case. Despite the time delay, the progression of each flare is remarkably similar at both frequencies.  Figure~\ref{fig:jt} shows normalised flare profiles for the 39.2 km s$^{-1}$ feature at both frequencies.  The normalisation was done by finding the maxima and minima for each time series, subtracting the minimum from each data point and then dividing by the difference between the maxima and minima. This is strong evidence for the cause of the flare being external to the masing region.

\begin{table*}
\begin{center}
\caption{Statistics for time series and results of period search. Columns marked 'old' refer to data taken between January 1999 and April 2003 and columns marked 'new' refer to data taken between August 2003 to November 2007.}
\label{tab:stats}
\begin{tabular}{lrrrrrrrrrrr}
\hline
	& \multicolumn{2}{c}{Average Flux Density} & \multicolumn{2}{c}{S/N}		& \multicolumn{3}{c}{Most sig. P (epochfold)}&\multicolumn{3}{c}{Most sig. P (DFT)}\\
& \multicolumn{2}{c}{(Jy)} & \multicolumn{2}{c}{}		& \multicolumn{3}{c}{(days)}&\multicolumn{3}{c}{(days)}\\	
		&	old	&	new	&	old	&	new	&	old	&	new	& 	all	&	old	&	new	&	all	\\
\hline
12.2 GHz, 39.25 km/s	&	30.9	&	39.5	&	15	&	120	&	29.51	&	29.40	& 	29.53	&	29.65	&	29.48	&	29.65	\\
6.7 GHz, 30.18 km/s	&	3.9	&	1.6	&	8	&	5	&	--	&	--	& 	--	&	--	&	--	&	--	\\
6.7 GHz, 31.74 km/s	&	3.8	&	4.6	&	8	&	13	&	29.60	&	29.43	& 	29.56	&	29.48	&	29.48	&	29.65	\\
6.7 GHz, 31.85 km/s	&	3.8	&	4.3	&	8	&	12	&	29.58	&	29.41	& 	29.53	&	29.65	&	29.48	&	29.48	\\
6.7 GHz, 32.75 km/s	&	8.1	&	9.8	&	17	&	28	&	29.33	&	29.44	& 	29.53	&	29.48	&	29.48	&	29.65	\\
6.7 GHz, 33.43 km/s	&	15.9	&	15.9	&	33	&	45	&	29.56	&	--      	& 	--	&	29.65	&	29.32	&	--	\\
6.7 GHz, 33.65 km/s	&	44.0	&	53.2	&	92	&	152	&	29.54	&	29.39	& 	29.51	&	29.65	&	29.32	&	29.48	\\
6.7 GHz, 34.33 km/s	&	1.4	&	2.0	&	3	&	6	&	--	&	--	& 	--	&	--	&	--	&	--	\\
6.7 GHz, 34.66 km/s	&	7.6	&	4.4	&	16	&	13	&	29.78	&	29.50	& 	29.56	&	29.65	&	29.48	&	29.65	\\
6.7 GHz, 36.57 km/s	&	3.6	&	4.1	&	8	&	12	&	29.53	&	29.29	& 	29.53	&	29.65	&	29.48	&	29.65	\\
6.7 GHz, 37.81 km/s	&	6.7	&	11.2	&	14	&	32	&	29.59	&	29.49	& 	29.57	&	29.65	&	29.48	&	29.48	\\
6.7 GHz, 38.71 km/s	&	24.9	&	17.5	&	52	&	50	&	29.48	&	29.49	& 	29.56	&	29.65	&	29.48	&	29.65	\\
6.7 GHz, 39.16 km/s	&	49.3	&	75.5	&	103	&	216	&	29.54	&	29.49	& 	29.55	&	29.65	&	29.48	&	29.48	\\
\hline
\end{tabular}
\end{center}
\end{table*}

\begin{table}
\begin{center}
\caption{Results of search for time lags relative to the 39.16 km s$^{-1}$ feature at 6.7 GHz.  A negative sign means the peak in question flares before the reference.}
\label{tab:dcf}
\begin{tabular}{lrrr}
\hline
2nd peak	&delay 		& z-dcf &$\sigma_{z-dcf}$ \\
(km s$^{-1}$)	&	(days)	&	&	\\
\hline
39.25 (12 GHz)	&	-1.1	& 0.75	& 0.06\\
30.17		&	--	& --	& 0.15	\\
31.74		&	-2.0	& 0.63	& 0.11	\\
31.85		&	-2.7	& 0.31	& 0.12	\\
32.75		&	0.7	& 0.75	& 0.06	\\
33.43		&	0.3	& 0.73	& 0.11 \\
33.65		&	1.0	& 0.53  & 0.11\\
34.33		&	--	& --	& 0.14\\
34.66		&	1.7	& 0.55	& 0.13\\
36.57		&	-3.7	& 0.63	& 0.12	\\
37.81		&	-3.3	& 0.73	& 0.07\\
38.71		&	-1.7	& 0.82	& 0.08\\
\hline
\end{tabular}
\end{center}
\end{table}

\section{Discussion}
The variations observed may be classified as quasi-periodic because of irregularities in the light curves, but the underlying cause may very well be strictly periodic. The persistence of the phenomenon is remarkable - our data cover 110 cycles counting from the first observations in January 1999. The short period raises some questions about the cause of the periodicity - if it is a binary object the orbits are very close.  Salient points for consideration of the cause of the periodicity are:
\begin{itemize}
 \item Stable period over 110 cycles.
 \item Delays of up to 5.7 days ($>10\%$ of the period) between features.
 \item Varying amplitude of flares, phase of peak seems to vary.
 \item Minima occur at the same phase in the cycle.
 \item Similar flare profiles at 6.7 and 12.2 GHz despite a time delay.
\end{itemize}

The rapid variability seen in this source raises the question of scintillation. Intraday variability has been seen in Galactic OH masers and intrepreted as due to scintillation \citep{Cle91}. Scintillation has been observed in H$_2$O megamasers \citep[e.g.][]{2005AJ....129.1231M}. \citet{2003ApJ...585..653B} observed rapid quasi-periodic variations at two frequencies as well as annual variations in the quasar PKS1257-326. Our shortest sampling interval is six hours, during which no significant variation (other than the instrumental effect discussed earlier) is seen.  There is no annual variation, other than that caused by slight increases in system temperature during the rainy season. Furthermore, the time scales of scintillation are frequency-dependent \citep{1998MNRAS.294..307W}.  In the case of G12.89+0.49 we see that the variations at 6.7 and 12.2 GHz have the same time-scales. Thus it seems unlikely that the variations seen in this source are due to scintillation.

The nature of the periodicity needs to be considered in the light of the full range of periods observed in class II methanol masers.  This is the shortest period observed, while six other objects have periods in the range 133 to 505 days \citep{2008ASPC..387..124G}. Both stellar pulsations and stellar rotation for main-sequence stars are expected to be of the order of a few hours to  a few days. The only plausible mechanism that can give rise to such a large range of periods is a binary system. While the effects of binarity on disc accretion in massive stars have not been studied in great detail as yet (the simulations of \citet{2007ApJ...656..959K} show the formation of multiple cores, with a resolution down to 10 AU), there is growing observational evidence of massive binary systems during the formation phase. \citet{2006A&A...457..553C} report the NIR detection of a visual binary with components of similar brightness and a projected separation of 1200 AU. The system is associated with a compact HII region and molecular outflow.
\citet{2007ApJ...655..484A} conducted a radial velocity survey of infrared lines towards nine targets. Two of  the targets showed large radial velocity variations indicative of close massive binaries. 

The recent high resolution study of \citet{2008ApJ...673L..55B} of G12.89+0.49 is of interest in this discussion. They trace a warm disc with a velocity gradient. The disc radius (assuming a distance of 3.6 kpc) is $\sim$3600 AU.  The rotation profile exhibits deviation from a Keplerian profile, which they speculate could be either a gravitationally unstable, infalling, rotating toroidal envelope or a massive, self-gravitating disc. They estimate that the central protostar has a mass in the range 15 -- 20 M$_\odot$. Assuming Keplerian rotation and a circular orbit, for an orbiting body with mass much less than that of the primary the orbital radius would be 0.46 AU for a combined mass of 15 M$_\odot$, to 0.51 AU for the upper mass limit of 20  M$_\odot$.  This may not be unreasonable since \citet{2007ApJ...655..484A} report a massive binary system with a primary of 15 M$_\odot$, secondary of 9.5 M$_\odot$ and an orbital radius $\lesssim$ 0.4 AU. For G12.89+0.49, at 3.6 kpc the maximum angular separation  would be a few milli-arcsec -- detection of a binary system would require a resolution that currently cannot be achieved. Thus the lack of evidence of multiplicity in the highest resolution observations to date does not rule out the possibility of a close binary system.  Higher angular resolution observations to determine the kinematics of inner regions of the disc will be extremely valuable in interpreting the maser variability.

Given the presence of a binary system and an accretion disc, there are many scenarios that could give rise to periodic events. A binary star system can have up to three discs - a disc around each star as well as a circumbinary disc. Without knowing the geometry of this system we can only discuss low mass cases where periodically modulated light curves have been observed. An intriguing low mass case of periodic accretion has been reported by \citet{2007AJ....134..241J}. The UZ Tau E system is a spectroscopic binary with a period of $\sim$ 19 days with periodic variations in the \textit{BVRI} light curves. Some parallels with the behaviour of G12.89+0.49 can be noted: the light curve is non-sinusoidal, with the bright state lasting 60\% of the duty cycle, the event does not necessarily occur during every orbit, there are long-term trends and short-term scatter seen in addition to to periodicity. The behaviour is explained by a variable accretion rate, giving rise to flares.  Material can flow from the circumbinary disc to the circumstellar disc with a rate that is dependent on the phase of the orbit \citep{1996ApJ...467L..77A,2002A&A...387..550G}. The material could undergo a shock when it collides with the circumstellar disc. Enhanced accretion can also occur near periastron.  This depends on the eccentricity of the orbit.

Another group reported recurrent eclipses of a low mass pre-main-sequence star in the young cluster IC348 with a 4.7 year period \citep{2006ApJ...646L.151N}. The eclipse lasted for 3.5 years (75\%) of the cycle.  They concluded that orbital motion is the only plausible explanation for such a long period, but they could be seeing a single star that is only directly visible through a gap in a circumstellar disc or a binary star system with an eccentric orbit inclined to the plane of the circumbinary disc.  In the latter case one member of the binary system is obscured during a portion of the cycle.

We have established during the study of another periodic object that the maser region itself is not physically affected during a flare \citep{2005MNRAS.356..839G}. The persistence of the spectral features also points to an event external to the maser region. The long term variations in the maser intensities are most likely due to changing local conditions. \citet{Dur01} have speculated that suitable conditions for methanol masers can occur in the outer (more than 10000 AU) regions of massive discs in gravitational instabilities. There are too many unknowns at this stage to make it feasible to model the maser flares.  The flares could be originating in the radio continuum emission from the HII region, which is then amplified by the masers. The other option is that the infrared flux is changing, which would modulate dust temperatures and thus the pump rate for the masers. The propagation of radiation through the intervening circumstellar material would have to be modelled taking into account the geometry of the extended disc-like structure, which is poorly known at this stage. Detection of periodicity in a tracer other than the masers is necessary to reduce the number of variables in this situation.  The sub-mm seems the most promising, particularly with the wealth of thermal lines that have been observed by \citet{2007A&A...466..215L}. 

The delays and shifts in the central velocity of the spectral features are also worthy of further investigation.  The observations of \citet{Wal98} did not resolve the maser features at all. Higher resolution maps, preferably repeated over the course of a flare will help clarify whether there are real variations in the maser velocity or if we are simply seeing a less saturated maser flaring more strongly than its neighbours.

\section{Summary and conclusion}

The methanol maser source G12.89+0.49 undergoes quasi-periodic variations at 6.7 and 12.2 GHz with a period of 29.5$\pm$0.1 days.  Significant deviations are seen from cycle to cycle but the underlying trigger may be strictly periodic. Interstellar scintillation is unlikely to be present because both frequencies show the same time scales, hence the variation is intrinsic to the source. 

Time delays of up to 5.7 days are seen between spectral features at 6.7 GHz and a delay of 1.1 day is seen between the corresponding 12.2 and 6.7 GHz spectral features. These delays could be used to construct the three-dimensional structure of the masers if higher resolution maps were available. VLBI mapping to find the detailed maser distribution at both frequencies, the relative positions of the 6.7 and 12.2 GHz masers and a confirmation of the distance through parallax measurements would be very fruitful.

The stability of the period is best explained by a binary system. This source is an ideal test case for massive star formation scenarios. It is known that a hot core, outflow and a disc-like structure are present. Multi-wavelength studies of this source over the course of a flare cycle should yield a wealth of information that will help us interpret the maser variability and constrain maser models.  

\bibliographystyle{mn2e}
\bibliography{./refs-adsabs}

\label{lastpage}
\end{document}